# Bi-Sampling Approach to Classify Music Mood leveraging Raga-Rasa Association in Indian Classical Music

Mohan Rao B C[*], Vinayak Arkachaari[*], Harsha M N[*], Sushmitha M N[*], Gayathri Ramesh K K[*], Ullas M S[*], Pathi Mohan Rao[*], Sudha G[*], Narayana Darapaneni[#]

[*] Dept of Computer Science & Engineering, PES University, Bengaluru, India.  [#] Northwestern University SPS and Great Learning, darapaneni@gmail.com

*Abstract*—The impact of Music on the mood or emotion of the listener is a well-researched area in human psychology and behavioral science. In Indian classical music, ragas are the melodic structure that defines the various styles and forms of the music. Each raga has been found to evoke a specific emotion in the listener. With the advent of advanced capabilities of audio signal processing and the application of machine learning, the demand for intelligent music classifiers and recommenders has received increased attention, especially in the "Music as a service" cloud applications

This paper explores a novel framework to leverage the raga-rasa association in Indian classical Music to build an intelligent classifier and its application in music recommendation system based on user's current mood and the mood they aspire to be in

*Keywords—Music Recommender, Indian Classical Music, Raga, Audio Signal Processing, Music Classification Algorithm*

## I. INTRODUCTION

Indian classical music (Hindustani and Carnatic) is a rich tradition that originated in South Asia and can now be found worldwide. Indian classical music has developed into a nuanced, glorious art form over the centuries. Indian classical music, through a variety of melodic entities (ragas), ornamentation of notes, and rhythmic patterns, tries to unite the performer and the listener in the experience of emotions or bhava [1]. In Indian classical music, ragas constitute specific combinations of tonic intervals capable of evoking distinct emotions [2].

Indian Classical Music can be broadly classified into Hindustani Classical Music and Carnatic Classical Music. There are many similarities and differences between the two music traditions. One of the major differences is in the way "ragas" are classified. In Hindustani classical music tradition, ragas are classified based on Raga-Ragani and Thaat method while in Carnatic Music, ragas are classified based on Janaka and Janya method [3]. A deeper understanding of Indian classical music is out of scope for the present work, but the understanding that ragas evoke emotions in the listener is critical and is a well-researched area

The impact of music on mood and mental conditions is a well-researched area in human psychology.  A research study on the role of music therapy with the application of Indian classical music (specifically Darbari Kanada raga) has shown conclusive evidence that listening to Indian Classical Music lowered depression, anxiety, and stress among elderly adults [4]. Another research study concluded that the evidence for the beneficial effects of music on reward, motivation, pleasure, stress, arousal, immunity is mounting and is promising, yet preliminary [5]. Many studies have provided evidence that music has an impact on mood and specifically different genres evoke different emotions in the listener.  A study to ascertain the impact of classical, pop, and metallic music on the mood of the listener found statistically significant evidence in favor of the hypothesis that music would affect mood [6]

The recent advances in the field of audio signal feature extraction and automatic audio classification using machine learning algorithms have increased the research interest in the audio analysis field. Various methods have been proposed to extract features from audio/music files which would help in projecting the audio into numerical dimensions. Once the numerical features are available, the classic machine learning algorithms could be applied to build classifiers and recommendation systems. In a review paper, the researchers have consolidated features extraction techniques and classified them into 2 major groups – Physical audio features extraction techniques and Perceptual audio features extraction techniques [7]. There are much research works exploring the application of different machine learning algorithms to classify music genres by extracting features from the audio/music files. Depending on the application in terms of speech recognition, sound classification, or music genre identification various algorithms have been found performing well. In one of the research studies, researchers found that Support Vector Machines provided the best classification accuracy with a western music dataset with 10 classes [8]. Research on the application of machine learning algorithms to classify and detect raga in Indian classical music is not new. The researchers after a comparative study done with a data set having 14 ragas, after testing both classical Machine Learning and advanced deep learning models found that extreme gradient boosting algorithm performed best [9]

Backed with the knowledge of previous works and recent advancement in below mentioned 3 fields
1. Music & Human Psychology
2. Indian Classical Music & Emotions
3. Music/Audio Analytics & Data Science

The current objective is to Develop a music recommendation system to elevate the user experience and emotion using Machine Learning or Deep neural network algorithms trained on the Indian classical music leveraging rasa-raga relationship including contemporary music which is having close association with the particular raga/rasa

## II. LITERATURE SURVEY

The relationship between music, the mood the music inherently has, and the mood that it evokes in the listener has been of interest for researchers for a long and a tremendous body of work has been done over the years. In this section, an attempt is made to revisit the previous works that, either closely resemble the current work or provide the foundational knowledge necessary for the current objective.

"Human emotions" is a field of study which is both fascinating and difficult. Historically, the generally accepted theory was hard, if not impossible to study music emotions scientifically [11]. Peretz observed that it was generally been considered that emotions are too personal, elusive, and variable. The research study on the facial expression of emotions provided evidence that made the scientific community take the universality of association between discrete emotions and facial muscular patterns [12]. Armed with these insights, one might be reasonably sure of the hypothesis that the association of music and emotions is universal. The next frontier where research is going on still is in the measurement of emotions. The pre-requisite on this frontier is to bring objectivity given both music and emotions are subjective in nature. One of the earliest attempts was by Weld H. P. [17] in 1912, where an attempt was made to measure emotions/mood by observing the listener's physiological changes in terms of respiration, hear-beat, and distribution of blood supply (quantitative measurement) and reviewing the introspections of the listeners after listening (Qualitative Measurement). The more successful attempt was made by Hevner, K [18] in which the listeners were provided with 8 groups of adjectives with 6-11 adjectives in each group, which they would have to tick after listening to the music. This approach made the task of associating emotion to music more objective. Building on these proposed models, Zentner M et al. [19] conducted a series of experiments and studies resulting in the proposal of Geneva Emotional Music Scales (GEMS) with 9 scales and 45 emotion labels. But all of these methods fall under a category called standard self-report instruments. From the current work perspective, the insights from these studies are that generally, one needs to select a scale for emotion/mood measurement and conduct an experimental study or a survey to assign emotion/mood labels to songs/music. This is an expensive and time-consuming proposition and the solution would be to leverage a music tradition that already has established a structured association between music and moods. That is where we shift our focus to Indian Classical Music

According to the Carnatic Music – Book 1, Secondary Level Course - Theory Text Book, Indian classical music, through a variety of melodic entities (ragas), ornamentation of notes and rhythmic patterns, tries to unite the performer and the listener in the experience of emotions or bhava [13]. The research by A. Mathur e*t al.* [14], provides supporting evidence to the hypothesis that Indian classical music ragas evoke different emotions in the listener. Furthermore, evidence of therapeutic application of Indian classical music (specifically Darbari Kanada raga) has been found in the works of S. Chattergee and R Mukharjee [15], indicating Indian classical music not only evokes emotions, but also helps in reducing Stress, Anxiety, and Depression. The behavioral study using Carnatic Indian classical Music by G K Koduri and B Indurkhya [20], based on 750 responses from 48 people, observed that "we can certainly say that the raga encapsulates the audio patterns responsible for eliciting specific emotions". Tough the researchers found the evidence, though encouraging, was not conclusive in supporting the raga-rasa association [20]. There is reasonable evidence supporting the hypothesis that we could leverage the time-tested raga-rasa association in Indian classical music and specifically Carnatic Indian classical Music, to overcome the necessity of a survey of listers and their perception of music and mood association. If we can identify the raga in a music/song, we can associate a dominant mood with reasonable confidence. The only aspect that needs attention now is to understand how we can extract numerical features from a music/song and use the features to fit a classifier algorithm. The results of the classifier algorithm could be used to extract the probability of a music/song is closer to a rasa (moods) and use it as input to a mood-based music recommendation engine

With the advancement of machine learning and deep learning fields in the last 3 decades, there has been a lot of experiments conducted both in finding effective ways of extracting numerical features from audio files and also in finding the best performing classification algorithms for audio classification. This niche field is aptly named music information retrieval systems. In a comparative study by Tao Li e*t al.* [21], the researchers listed the content-based acoustic features extraction methods from audio, which includes Mel Frequency Cepstral Coefficients (MFCC), and proposed a new method called Daubechies Wavelet Coefficient Histograms (DWCH). The authors note that traditional feature extraction methods, more or less, capture incomplete information of music signals, thus establishing the need for a new method proposal. The researchers also compared the performance of various machine learning algorithms and feature extraction methods and concluded that the Support Vector Machines (SVM) algorithm with DWCH features performed better than other algorithms and features extraction combinations. F. Alias et al. [22], provided a comprehensive review of the most relevant audio techniques (as of 2016) where the techniques were organized into 2 broad categories (Physical and Perceptual) and then second-level classification into 7 types based on domain. The research placed Mel-Frequency Cepstral Coefficients (MFCC) under the Wavelet-Based Perceptual Features category of audio feature extraction techniques. The techniques of feature extraction from music have continued to evolve. K Choi conducted a comparative analysis between classical MFCC feature extraction technique with pre-trained convnet feature on a set of unrelated music regression and classification tasks and found that the convnet feature outperforms the baseline MFCC feature in all the considered tasks [26], providing a basis and justification for success of the application of transfer learning concept on music classification tasks

D. Liu et al. [23], demonstrated a hierarchical and non-hierarchical framework, using Intensity, Timbre, Pitch, and Rhythm features, to first detect and assign moods to music clips (using Gaussian Mixture Models). The method gave a reasonably good framework for mood detection in audio files, but it had only 4 classes of moods. S P.Sumare and D G Bhalke leveraged this framework and used MFCC features and an SVM algorithm to train a classifier to detect moods in music clips automatically in an IOSR journal paper in 2015 [24]. Seo, Yeong-Seok & Huh, Jun-Ho [25] demonstrated a different approach where they combined emotion survey data and extracted music feature data with a high correlation between reported emotions in the survey and the emotion prediction from SVM based classification model. This forms acceptable evidence that perceived emotions can be predicted with music features with confidence. The final question to be answered from the perspective of current work is how effective are the Music feature extraction techniques and classification algorithms for mood detection and prediction when applied to Indian Classical Music

The application of Music Information retrieval techniques and classification algorithms is not new and many successful attempts have been made in this direction. It is critical to review as most of the initial applications of machine learning algorithms and feature extraction techniques were developed on western music and non-music audio signal processing. In their 2007 paper, M S Sinith and K Rajeev experimented with the application of hybrid of Hidden Markov Model and Dynamic Time Warping for identification of 9 Carnatic Indian Classical Music ragas in monophonic audio files which had promising results which were better than the previous attempt by the same researchers using HMM models [27]. The task here was to identify raga only and not for emotion detection or classification. V Kumar e*t al.* [28], proposed a framework of extracting music features from 2 kernels, Pitch-class profiles, and n-gram distribution, for identification of Carnatic Indian Classical Music, and reported an improvement of 10% accuracy improvement compared to conventional pitch-class profile features. Sujeet Kini, Gulati, S, and Rao, P demonstrated that by applying timbral features, tempo, and modulation spectra of timbral features, classification of the sub-genres of north Indian music (bhajans and qawwali) at an accuracy of 92% could be achieved. The researchers observed that the misclassification errors in bhajans, misclassified as qawwalis by timbral features, corresponded generally to female singers. For qawwalis misclassified as bhajans, it was observed that vigorous drumming and handclapping were absent, and so was the stressed singing style and heavy chorus [30]. These insights help us understand the variety of songs/music to be covered in the data collection stage for the current work and the data cleaning that might be applied to improve classification accuracy. The closest research study to the current work was found in an IEEE paper submitted by P S Lokhande and B S Tiple titled "A Framework for Emotion Identification in Music: Deep Learning Approach" [28]. In this paper, the researchers have completed a survey of performances of different classification algorithms for music mood classification. The survey covered results reported from previous studies covering Multiple Linear Regression, Support Vector Regression, Simple Neural Network, Naïve Bayes, Support Vector Machines, Hidden Markov Models, Deep Belief Networks, and an ensemble of DBN and HMM. Researchers have leveraged the raga-rasa association in Indian Classical Music. The researchers did not report the results but have concluded that Deep Belief Network provided good results. It is also unclear in the paper whether the researchers' used songs in pure classical vocal rendition or if movie songs were used as input data to train the models. But in the conclusion, researchers mentioned the application to be in music therapy and psychological treatments, which hints at songs with pure classical reeditions.

Our current work is unique in the sense that the application aimed at is to propose a mood-based recommendation system. The input data used for training the classifier is also unique in that the sample has songs from both movies and classical renditions covering 9 languages and includes a few instrumental renditions of the raga also. The labeling of the songs into ragas into rasa has been done in consultation with Indian Classical Music Practitioner and the data labels have been validated extensively. The work is also unique in terms of a sampling technique where 2 samples at different time intervals have been considered

## III. EXPERIMENTAL SETUP

### A. High-level plan

The below diagram (fig. 1) represents the overall experimental setup used to

1. Ascertain the Rasa (mood) association with ragas in Indian classical music (Hindustani & Carnatic) through research into Indian Classical Music and in consultation with domain expert
2. Collect the songs (mix of classical renditions and Indian Movie songs) for each of the ragas and associate rasa (mood)
3. Build a classifier model on collected songs to predict rasas (moods)
4. Use the classifier model to classify new songs/music to be used in the recommendation system

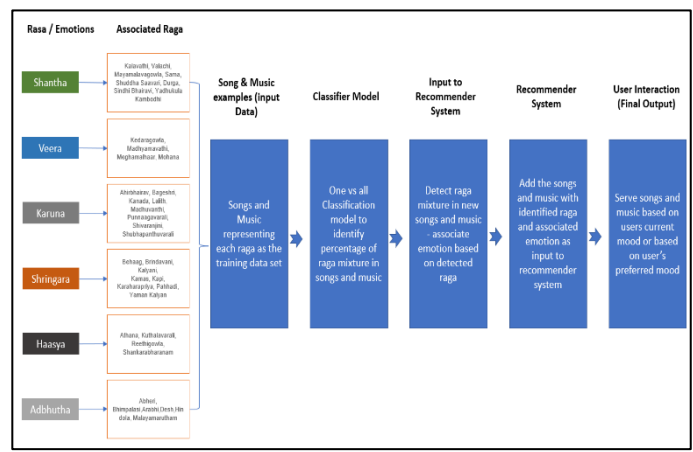

Fig. 1 - Representation of the overall flow of the experiment and expected output

In general, the accepted number of rasa in Indian classical music and Indian drama/acting is 9 (Shringara - love/beauty), Hasya - laughter, Karuna - sorrow, Raudra - anger, Veera - heroism/courage, Bhayanaka - terror/fear, Bibhatsya - disgust, Adbutha - surprise/wonder, and Shantha - peace or tranquillity), But music is not generally rendered to evoke emotions like "Bhibhatsa" (disgust), "Bhayanaka" (fear/anxiousness), and "Roudra" (anger). So, it would be hard to find the songs with raga associated with these emotions and for that reason, these 3 moods are excluded from the scope of this work.

The raga/rasa association finalized after review of musician blog posts and validation in consultation with Indian Classical Music Domain expert is as shown below the table

TABLE 1 - Raga/Rasa Association Matrix – Based on Indian Classical Music Domain Expert Knowledge & opinion

| *Adhbhutha* | *Haasya* | *Karuna* |
|---|---|---|
| Abheri/Bhimpalasi | Aathana | Ahibhairav |
| Arabhi | Kunthalavarali | Bageshri |
| Desh | Reethigowla | kanada |
| Hindola | Shankarabharanam | Lalith |
| Malayamarutham | | madhuvanti |
| | | Punnagavarali |
| | | Shivaranjani |
| | | ShubhapanthuvaraLi |

| *Shantha* | *Shringara* | *Veera* |
|---|---|---|
| Kalavathi/valachi | Behaag | Kedaragowla |
| Mayamalavagowla | Brindavani | Madhyamavathi |
| Sama | Kalyani | Meghamalhaar |
| Shuddha Saveri - Durga | Kamas | Mohana |
| Sindhu Bhairavi | Kapi | |
| Yadhukula kambhodhi | Karaharapriya | |
| | Pahaadi | |
| | YamanKalyani | |

## IV. Data Collection

A pre-requisite for the data collection is to derive a data collection strategy based on the core objective of the current work and data availability. A study on cross-cultural similarities and differences of music-evoked emotions, including test subjects from India and Finland, by S. Saarikallio e*t al.* [16], noted that the most prevalent music genre for a mood management function in Indians was film music. This indicates the sample data should have Indian movie songs with significant representation. Given that the current work is focused on the Carnatic class of Indian classical music, which is predominant in the southern part of India, the movie songs from the southern part of India (includes Kannada, Tamil, Telugu, and Malayalam language movie songs). Another critical requirement is to make sure the classification model also can learn from the pure classical reeditions of the raga, either vocal or instrumental. So, the sample data needs to have pure classical renditions representation also. To further expand the population from which the sample data would be selected, we could look at Bollywood songs and music also. To achieve this, we could leverage the relationship with a few Carnatic ragas with Hindustani ragas in terms of being close in structure and aesthetics. This would help us include Hindi songs both movie songs and classical renditions

The songs were identified to be belonging to a raga category by looking at the playlists created by users on YouTube and other music platforms. The blog posts of the musicians and vocal singers were reviewed to find as many songs and music examples as could be collected. Songs downloaded in WAV format from YouTube for each raga and rasa as the python library used for extracting features was Librosa and this library works well with audio files in WAV format. The total sample size is as shown in table 2 below

TABLE 2 - Input Data Samples per Rasa/Mood

| *Rasa/Mood* | *# Of Song/Music files* |
|---|---|
| Adhbhutha | 137 |
| Haasya | 79 |
| Karuna | 119 |
| Shantha | 119 |
| Shringara | 216 |
| Veera | 146 |
| **Grand Total** | **816** |

TABLE 3 - Input Data Samples per genre

| *Genre* | *# Of Song/Music files* | *% Of Song/Music files* |
|---|---|---|
| Folk/Album | 9 | 1.10% |
| Indian Classical | 104 | 12.75% |
| Movie | 703 | 86.15% |
| **Grand Total** | **816** | **100.00%** |

TABLE 4 - Input Data Samples per Language

| *Language* | *# Of Song/Music files* | *% Of Song/Music files* |
|---|---|---|
| Tamil | 324 | 39.71% |
| Hindi | 221 | 27.08% |
| Kannada | 102 | 12.50% |
| Telugu | 66 | 8.09% |
| Malayalam | 60 | 7.35% |
| Sanskrit | 29 | 3.55% |
| Instrumental | 12 | 1.47% |
| English | 1 | 0.12% |
| Marathi | 1 | 0.12% |
| **Grand Total** | **816** | **100.00%** |

*A. Challenges with the data collection*

- Acquiring songs labeled by Raga is challenging. Only an expert for the Indian Carnatic music can identify raga
- The available songs are not always in clean shape. It would have intro speech, clapping, a mixture of music, etc, which might need treatment
- Not all ragas are employed in movie songs, so finding data for all ragas is challenging

*B. Overcoming the data collection challenges*

- Reviewed blog posts of Indian Music enthusiasts and artists to identify the songs and music under each raga
- Reviewed the playlists built by users in YouTube to download the songs belonging to a particular raga
- Reviewing the methods of data cleaning using audio tools available online
- Limited the scope of ragas where the songs are not available and also limiting the mood/Rasa to the ones where songs are available

## V. Audio Feature Extraction & EDA

The first step in the process of building a music mood classifier would be to transform the audio wav files to numerical features to feed the data into a classifier algorithm. There are many methods available to extract features out of audio files and each method has been found effective in different scenarios. Sound features can be used to detect speakers, detect gender, age, diseases, and much more through the voice [10]. The general process is to break down the audio file into windows of equal size and then extract the features of each window. Some of the most used audio feature extraction methods are listed below with a short description

*1) Statistical Features* - mean, standard deviation, minimum, maximum, median, and quartiles of the frequencies of each signal for each window [10]

*2) Energy* - The energy of a signal is the total magnitude of the signal, i.e. how loud the signal is. [10]

*3) Root Mean Square Energy* - The RMS Energy (RMSE) is simply the square root of the mean squared amplitude over a time window [10]

*4) Zero-Crossing Rate* - The zero-crossing rate indicates the number of times that a signal crosses the horizontal axis, i.e. the number of times that the amplitude reaches 0 [10]

*5) Tempo* - An estimate of the tempo in Beats Per Minute (BPM) [10]

*6) Mel Frequency Cepstral Coefficients (MFCC)* - The Mel-frequency cepstrum (MFC) is a representation of the short-term power spectrum of a sound, based on a linear cosine transform of a log power spectrum on a nonlinear Mel scale of frequency [10]. Mel-frequency cepstral coefficients (MFCCs) are coefficients that collectively make up an MFC [10]. They are derived from a type of cepstral representation of the audio clip (a nonlinear "spectrum-of-a-spectrum")

For the current research work, MFCC has been adopted to extract features from the input audio files. In the paper titled "Mel frequency cepstral coefficients for Music Modelling", author (Beth Logan) had the following conclusion to make, regarding the application of MFCC in music modeling "We found that the Mel scale was at least not harmful for this problem, although further experimentation is needed to verify that this is the optimal scale for modeling music spectra in the general case" [11]

A. *Mel Frequency Cepstral Coefficients (MFCC) as the Feature Extraction Method*

The MFCC is a representation defined as the real cepstrum of a windowed short-time signal derived from the fast Fourier transform of the speech signal. The difference from the real cepstrum is that a nonlinear frequency scale is used, which approximates the behavior of the auditory system. The MFCC could be as follows [29]

Given that $x[n]$ denote the $N$ samples of a speech waveform, discrete Fourier transform (DFT) of the input speech signal is defined as follows

$$X[k] = \sum \sum_{n=0}^{N-1} x[n] e^{-j2\pi nk/N}$$

we define a filterbank with $M$ filters ( $m = 1,2,..,M$ ), where filter $m$ is triangular filter given by:

$$H[m,k] = 0 \qquad if \quad k < f[m-1]$$

$$H[m,k] = \frac{k = f[m-1]}{f[m] - f[m-1]} \quad if \quad f[m-1] \leq k \leq f[m]$$

$$H[m,k] = \frac{f[m+1] - k}{f[m+1] - f[m]} \quad if \quad f[m] \leq k \leq f[m+1]$$

$$H[m,k] = 0 \qquad if \quad k > f[m+1]$$

Such filters compute the average spectrum around each center frequency with increasing bandwidths, and they are displayed in Fig. 2

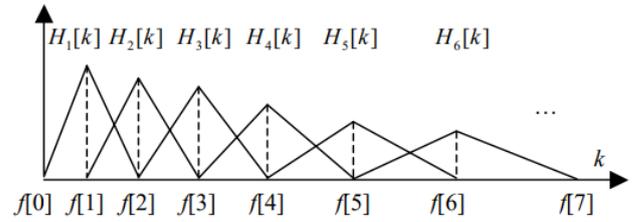

Fig. 2 - Triangular filters used in the computation of the mel-cepstrum

Let's define $f_l$ and $f_h$ to be the lowest and highest frequencies of the filter bank in Hz, $F_s$ the sampling frequency in Hz, $M$ the number of filters, and $N$ the size of the Fast Fourier Transform (FFT). The boundary points $f[m]$ are uniformly spaced in the mel-scale:

$$f[m] = \left(\frac{N}{F_s}\right) B^{-1}\left(B(f_l) + m \frac{B(f_h) - B(f_l)}{M+1}\right)$$

where $B(f) = 1125 \ln\left(1 + \frac{f}{700}\right)$

and $B^{-1}(b) = 700(e^{\left(\frac{b}{1125}\right)} - 1)$

We then compute the log-energy at the output of each filter as

$$S[m] = ln\left[\sum_{k=0}^{N-1} |X_a[k]|^2 H_m[k]\right], \qquad 0 \leq m < M$$

The MFCC is then the discrete cosine transform of the M filters outputs:

$$c(n) = \sum_{m=0}^{M-1} S[m] \cos\left(\frac{\pi n(m - 0.5)}{M}\right) \qquad 0 \leq n < M$$

where M varies for different implementations from 24 to 40. For speech recognition, typically only the first 13 cepstrum coefficients are used. It is important to note that the MFCC representation is no longer a homomorphic transformation [29].

MFCCs are commonly used as features in speech recognition systems, such as the systems which can automatically recognize numbers spoken into a telephone.

MFCCs are also increasingly finding uses in music information retrieval applications such as genre classification, audio similarity measures, etc

## B. Exploratory Data Analysis & Data Preparation

The first step in exploratory data analysis was to listen to each audio file and then clean it up by removing the portions of the music which were not relevant. Many of the music songs have speech, applause, other music, etc which were removed to clean up the audio file

The second step was to ascertain a standard length of audio/music file to be used to extract MFCC features as it is not necessary to have the full-length files as input to the classifier model. After consultation with an Indian classical music domain expert of the length of the song required to classify the song to the right raga, it was found that the first 60 seconds of the audio/music file should be a good candidate. This again is a hyperparameter and could be optimized based on classification accuracy we get

The third step was to conduct a visual analysis of the audio/music files from different mood/rasa classes to understand if there are clear visual differences between the representation of the audio/music files. For this purpose, 3 songs of 60-second length, belonging to the mood/rasa class of "Shringara" and "Adhbhutha", were chosen randomly. The visual representation of the songs in the form of Waveplot (Fig. 3) and MFCC Spectrogram (Fig. 4) was compared.

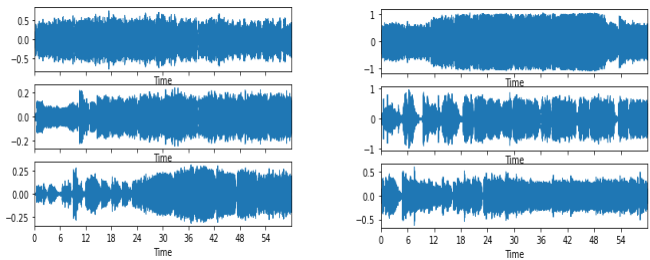

Fig. 3 – Waveplot of sample song files ("Adhbhutha" mood/rasa on the left and "Shringara" mood/rasa on the right)

Apart from the visible differences in the Waveplot range, which is generally between -1 to 1 for songs in the "Shringara" class and -0.5 to 0.5 for songs in the "Adhbhutha" class, no other distinguishing features could be observed

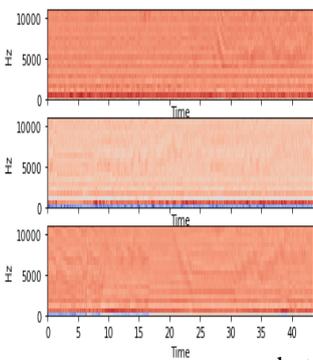

Fig. 4 – MFCC Spectrogram of sample song files ("Adhbhutha" mood/rasa on the left and "Shringara" mood/rasa on the right)

It was challenging to analyze the differences between the MFCC spectrograms with naked eyes, but there seem to be lesser intensity pixels for "Shringara" class songs compared to "Adhbhutha" class songs. Though the sample size is too low to draw any reliable conclusion, the plot helps in better understanding of the structural differences in the numerical representation of the songs from different classes

The fourth step was to extract MFCC features from each of the audio files and decide on no of features, which is a hyperparameter. This would be decided based on the accuracy of the classifier model and after a bit of experimentation, 40 features seemed to work well.

With these steps, the data is ready for a classification model with 40 MFCC features for each audio/music file and the label of the mood-based raga the song belongs to. The next step is to apply various classification algorithms to ascertain which algorithm works best and how it could be improved with hyperparameter tuning

Below is the correlation heat map of 40 features finalized for the classification model with audio/music file length of 60 seconds (0 to 60 seconds) segment (Fig. 5). Many of the features are correlated, but the accuracy of the classification model is the ultimate metric for model performance

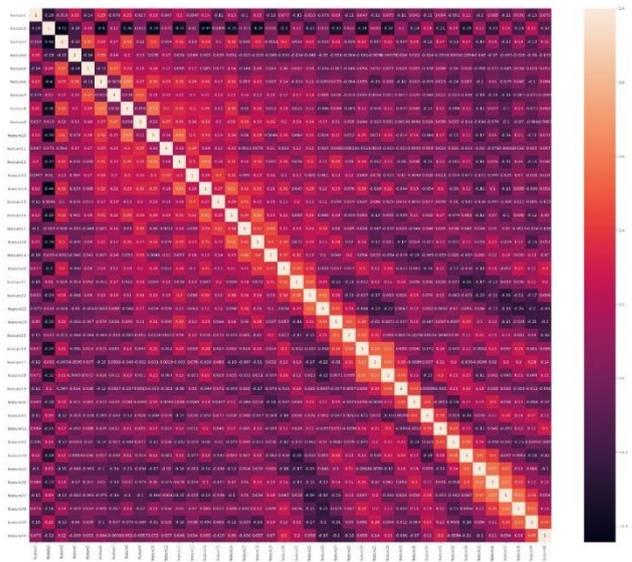

Fig. 5 - Input features correlation heat map (40 MFCC features per audio file)

## VI. OVERALL FLOW OF RESEARCH ACROSS STAGES

The overall research workflow and architecture follow a typical machine learning model development lifecycle. The detailed flow of tasks has been illustrated in the below flow chart (Fig. 6).

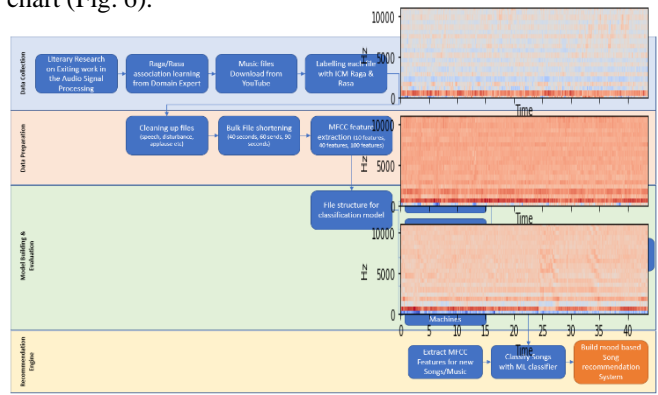

Fig. 6 - Detailed Steps and Project Overall Architecture

The classification algorithms chosen for building the classifier to predict the rasa/mood of the song are listed as follows

1) *Simple Artificial Neural Networks (ANN)*
2) *Naïve Bayes Classifier*
3) *K Nearest Neighbour Classifier (KNN)*
4) *Random Forest Classifier*
5) *Support Vector Machines Classifier (SVM)*

The early results from the first iteration of classifier model building and tuning the model for improved classification accuracy, are discussed in the next section

## VII. CLASSIFIER MODEL DEVELOPMENT

The current challenge can be categorized into a multiclass classification problem with audio/music MFFC features as input features and the rasa/mood as the output label. The data has 816 sample audio/music samples

### A. Base Models & Early Results

The base model was built on 4 variants of audio files as input samples

1) *Audio Files of 60 seconds length, covering the first 60 seconds of audio/music*
2) *Audio Files of 90 seconds length, covering the first 90 seconds of audio/music*
3) *Audio Files of 60 seconds length, skipping first 20 seconds and covering next 60 seconds of audio/music*
4) *Audio Files of 40 seconds length, skipping first 20 seconds and covering next 40 seconds of audio/music*

The results of the base classification model building on validation data set for various algorithms are discussed below

TABLE 5 – KNN Classifier 1st Iteration Results

| Sl No | Algorithm | Song Start Point | Song Duration | Validation Classification Accuracy | Model Architecture / Parameters |
|---|---|---|---|---|---|
| 1 | K Nearest Neighbours Classifier | 0 | 60 | 0.26 | Standard scaled data, Base Model with default parameters |
| 2 | | 0 | 60 | 0.27 | Standard scaled data, parameter tuning n_neighbors = 15, metric='manhattan' |
| 3 | | 0 | 90 | 0.18 | Standard scaled data, Base Model with default parameters |
| 6 | | 0 | 90 | 0.25 | Standard scaled data, parameter tuning n_neighbors = 9, metric='hamming' |
| 4 | | 20 | 40 | 0.27 | Standard scaled data, Base Model with default parameters |
| 7 | | 20 | 40 | 0.23 | Standard scaled data, parameter tuning n_neighbors = 19, metric='euclidean' |
| 5 | | 20 | 60 | 0.23 | Standard scaled data, Base Model with default parameters |
| 8 | | 20 | 60 | 0.12 | Standard scaled data, parameter tuning n_neighbors = 7, metric='hamming' |

TABLE 6 – Random Forest Classifier 1st Iteration Results

| Sl No | Algorithm | Song Start Point | Song Duration | Validation Classification Accuracy | Model Architecture / Parameters |
|---|---|---|---|---|---|
| 1 | Random Forest | 0 | 60 | 0.29 | Standard scaled data + base model |
| 2 | Random Forest | 20 | 60 | 0.35 | Standard scaled data + base model |
| 3 | Random Forest + Hyper Parameter Tuning | 20 | 60 | 0.33 | {'criterion': 'entropy', 'max_depth': 30} |
| 4 | Random Forest | 0 | 90 | 0.29 | Standard scaled data + base model |
| 5 | Random Forest + Hyper Parameter Tuning | 0 | 90 | 0.33 | {'criterion': 'entropy', 'max_depth': 70, 'max_features': 0.5, 'min_samples_leaf': 5, 'min_samples_split': 10} |
| 6 | Random Forest | 20 | 40 | 0.2 | Standard scaled data + base model |
| 7 | Random Forest + Hyper Parameter Tuning | 20 | 40 | 0.26 | {'criterion': 'entropy', 'max_depth': 50, 'max_features': 0.9, 'min_samples_leaf': 5, 'min_samples_split': 2, |

|   |   |   |   |   | 'n_estimators': 200} |
|---|---|---|---|---|---|

TABLE 7 – Support Vector Machines Classifier 1st Iteration Results

| Sl No | Algorithm | Song Start Point | Song Duration | Validation Classification Accuracy | Model Architecture / Parameters |
|---|---|---|---|---|---|
| 1 | SVM | 0 | 60 | 28.4 | Standard scaled data + base model |
| 2 | SVM + Hyperparameter Tuned | 0 | 60 | 31.8 | Standard scaled data + base model tuned parameters used : kernel='rbf',C=100,gamma=0.001 |
| 3 | SVM | 20 | 60 | 25.9 | Standard scaled data + base model |
| 4 | SVM + Hyper Parameter Tuned | 20 | 60 | 35.7 | Standard scaled data, {'C': 1, 'gamma': 0.1, 'kernel': 'rbf'} |
| 5 | SVM | 0 | 90 | 27.4 | Standard scaled data + base model |
| 6 | SVM + Hyper Parameter Tuned | 0 | 90 | 31.8 | Standard scaled data, {'C': 1, 'gamma': 0.01, 'kernel': 'rbf'} |
| 7 | SVM | 20 | 40 | 19.6 | Standard scaled data + base model |
| 8 | SVM + Hyper Parameter Tuned | 20 | 40 | 24 | Standard scaled data, kernel='rbf',C=10,gamma=0.001 |

TABLE 8 – Simple Artificial Neural Networks Classifier 1st Iteration Results

| Sl No | Algorithm | Song Start Point | Song Duration | Validation Classification Accuracy | Model Architecture / Parameters |
|---|---|---|---|---|---|
| 1 | Simple ANN | 0 | 60 | 0.2653 | 4 Hidden Layer, 100 Epochs, batch size 50 |
| 2 | Simple ANN | 20 | 60 | 0.2408 | 4 Hidden Layer, 100 Epochs, batch size 50 |
| 3 | Simple ANN | 20 | 40 | 0.249 | 4 Hidden Layer, 100 Epochs, batch size 50 |
| 4 | Simple ANN | 0 | 90 | 0.2735 | 4 Hidden Layer, 100 Epochs, batch size 50 |

*B. Insights from Initial Classification Model Trials*

   *1) Even with the variation of the audio/music file length the validation accuracy is not breaching 36%*

   *2) The models were tried with 10, 40, and 100 MFFC features per audio/music file, the accuracy did not improve*
   *3) Hyperparameter tuning with grid search also did not contribute to the validation data accuracy improvement*
   *4) The reasons might be the following*
      *a) The Sample data size for the model is low, needs to be increased*
      *b) The features are not able to differentiate, so other methods of feature extraction might have to be tried*
      *c) The audio files might have to be standardized for noise and other audio features*
   *5) Data processing could be augmented with PCA and running the model on the PCA features to see if the validation accuracy goes up*

*C. Classifier Model Improvements*

With all the learnings from the base model, the following are the strategies applied to improve the validation set accuracy

   *1) Apply scaling to the MFCC features before the model fitting*
   *2) Increase the sample size by taking 2 segments of the audio file, as the base model has only one segment of the audio file*
   *3) If the accuracy does not improve, then apply PCA to the input MFCC features for the data set with 2 segments of audio for each file and then fit the model on PCA features*

The scaling of input features did not help much in improving the accuracy. Then with the consultation of a classical music domain expert, a new strategy to increase the input data size was formulated. 2 segments of each audio file were considered for the input data set. The first segment was 60 seconds in length covering the first 60 seconds of the music file. The second segment skipped the first 20 seconds of the song and considered the next 60 seconds of the music file. This doubled the input data size from 816 to 1,632 and the results are as shown below

TABLE 9 – Model validation accuracy after considering 2 segments of each audio file as input for models

| Sl No | Algorithm | Song Start Point | Song Duration | Validation Classification Accuracy | Model Architecture / Parameters |
|---|---|---|---|---|---|
| 1 | SVM | 2 Segments - 0s-60s and 20s-80s | 120 | 0.73 | Standard scaled data + base model |
| 2 | SVM + Hyper Parameter Tuning | 2 Segments - 0s-60s and 20s-80s | 120 | 0.77 | kernel='rbf',C=10,gamma=0.1 |
| 3 | Random Forest | 2 Segments - 0s-60s | 120 | 0.67 | Standard scaled data + base model |
| | | and 20s-80s | | | |
| 4 | Random Forest + Hyper Parameter Tuning | 2 Segments - 0s-60s and 20s-80s | 120 | 0.69 | {'criterion': 'entropy', 'max_depth': 50, 'max_features': 0.5} |
| 5 | K Nearest Neighbours Classifier - Standard Scaling | 2 Segments - 0s-60s and 20s-80s | 120 | 0.84 | Standard scaled data, parameter tuning n_neighbors = 1, metric='manhattan' |
| 6 | K Nearest Neighbours Classifier - Standard Scaling | 2 Segments - 0s-60s and 20s-80s | 120 | 0.70 | Standard scaled data, parameter tuning - n_neighbors= 3 , weights = 'distance', metric='manhattan' |
| 7 | K Nearest Neighbours Classifier - Z Score Scaling | 2 Segments - 0s-60s and 20s-80s | 120 | 0.72 | Z score scaled data, parameter tuning - n_neighbors= 3 , weights = 'distance', metric='manhattan' |
| 8 | Simple ANN | 2 Segments - 0s-60s and 20s-80s | 120 | 0.46 | 4 Hidden Layer, 200 Epochs, batch size 10 |
| 9 | Logistic Regression | 2 Segments - 0s-60s and 20s-80s | 120 | 0.39 | Z score scaled data, max_iter=1000 |
| 10 | Naïve Bayes - Gaussian | 2 Segments - 0s-60s and 20s-80s | 120 | 0.36 | Z score scaled data, Default parameters |

*D. Classifier Model Final Selection*

Based on the validation accuracy results, the KNN classifier with n-neighbors = 1 has the best result. But this cannot be accepted as the neighbors considered for classification with 1 and it is not a robust model. The second-best results are obtained by the SVM model with parameter optimization with grid search with a validation accuracy at 0.77. This is the final model selected for the classification of songs into moods for the new songs

*E. Insights on Model Performance*

  *1)* Based on validation results, the final non-linear SVM model has been found weak in identifying the "Haasya" rasa (66%) while it was reasonably good in identifying "Shringara" rasa. This seems to correlate with the sample size as 26% of input sample data belonged to the "Sringara" class and 10% represented by the "Haasya" class – increasing the representation of "Haasya" and keeping the representation of all rasa equal might help in improving the accuracy

  *2)* A significant difference between training accuracy (100%) and validation accuracy (77%) indicates an element of overfitting of the model. This could also be treated by balancing the rasa classes representation in input data

  *3)* The model is equally accurate in predicting the rasa of Movie songs and pure Indian classical rendition but has lower accuracy in the case of folk songs. But the folk song represents 1% of the input population and increasing this representation might improve the folk song rasa prediction accuracy and in turn the overall accuracy

  *4)* Only one English song was part of input data and the model has predicted the rasa label accurately. The same is the case for Marathi songs also, There were 5 instrumental renditions in the validation data set and all of them have neem accurately classified into rasas, which is interesting to observe

  *5)* The general observation is that the error rates for different rasa classes are correlated to the sample representative of the class in input data and by balancing the classes in our data the overfitting could be reduced and classification accuracy could be improved

VIII. CONCLUSIONS AND FURTHER RESEARCH

The literature research conducted during this research study uncovered many previous works. But the scope of those research works could be classified into 3 categories

  *1) Exploring the relationship between Music & Human Psychology*

  *2) Exploring the relationship between Indian Classical music & human emotions*

  *3) Impact of different methods of Audio feature extraction and application of different classification algorithms on the classification accuracy with music as input data*

The Uniqueness of the present research is that it is an attempt at combining the learnings from the work done in the above 3 categories into a real-life application. The encouraging results from the application of classification accuracy prove the initial hypothesis that ML algorithms can identify the Rasa/emotional category of the songs through Indian Classical Music raga/rasa association intelligence to mimic the capability of an Indian Classical Music domain expert. Some of the previous attempts in the application of classification ML algorithms to predict the raga of the audio file were tested on pure raga renditions in their original form. But the current work is tested on movie songs adoption of the raga, which could also be considered as a first of the kind effort

The work paves the way to some interesting practical applications in music classification and music recommendation systems. Historically the songs were classified into different emotion/mood categories either manually or through a survey where responders listened to music and responded with a mood that music evokes based on perception. The other strategy would have been to overcome the cold start problem in the recommendation engine by providing rule-based recommendations and improving it with feedback from users. The current work helps overcome these challenges by leveraging the raga/rasa association in Indian Classical Music that could be applied to any music for automatic classification.

The current work is not devoid of shortcomings and challenges, which the researchers hope the future researchers would be able to resolve with further study. Accuracy of 0.77 on validation with SVM, though encouraging, has scope for improvements with the application of more sophisticated and advanced deep learning frameworks. More experimentation on the music file segmentation strategy may contribute to improvements in the accuracy of the classifier. One more avenue to be explored is combining the features from various feature extraction methods as input features for classification algorithm, which we leave on the able shoulders of future researchers

ACKNOWLEDGMENT

The authors thank Ullas M S for providing the guidance and bringing in the critical domain knowledge of Indian Classical Music. The authors also thank Sudha B G and Dr. Narayana Darapaneni for valuable inputs and support for completing the research work